\documentclass[11pt]{article}

\usepackage[round]{natbib}
\usepackage{a4}
\usepackage{color}
\usepackage{hyperref}
\usepackage{graphicx}
\usepackage{array, blindtext}
\usepackage{caption}
\usepackage{lscape}
\extrafloats{100}
\usepackage{tablefootnote}
\usepackage{multirow}
\def\GEOSauth#1 {\uppercase{ #1} \vskip 0.3cm plus 0.1cm minus 0.1cm}
\newcounter{authno}
\def\GEOSinst#1 {\small \addtocounter{authno}{1} $^{\arabic{authno}}$ #1
        \vskip 1mm \large}
\def\GEOSx#1 {\small \ts $^*$ #1 \vskip 1mm \large}
\def\GEOSinsto#1 {\small \ts #1 \vskip 1mm \large}

\title{\vspace{-2.5cm}
\large
{\bf GEOS RR 60  \hspace{2cm} GEOS CIRCULAR ON RR LYRAE  \hspace*{\fill}  February 1,2019} \\
\vspace{1.5cm}
\Large
{\uppercase {\bf Observations of the RRc variable LINEAR 1169665 \\
with the robotic telescope TAROT}}
\vspace{0.5cm}
}
\author{J.F. Le Borgne$^{1,2}$, A. Klotz$^{1,2}$}
\date{\vspace{-5ex}}    % remove the date in title using \maketitle

\begin{document}
\maketitle

\GEOSinst{GEOS (Groupe Europ\'een d'Observations Stellaires), 23 Parc de Levesville, 28300 Bailleau l'Ev\^eque, France}
\GEOSinst{IRAP; OMP; Universit\'e de Toulouse; 14, avenue Edouard Belin, F-31400 Toulouse, France}
\vspace{5mm}

\begin{abstract}
LINEAR 1169665 is a RR Lyrae of sub-type c discovered by the asteroid survey
LINEAR. The robotic telescope TAROT at Calern Observatory has observed it
between the years 2006 and 2015. \\
The present study of TAROT data as well as those from the surveys ASAS-SN
(2012-2019) and LINEAR (2002-2008) shows that this star presents a period
modulation with a period of 1800.1 days (4.9 years)
but with no significant variation of magnitude at maximum.\\
This phenomenon is similar to the one found in KEPLER data for the RRc KIC
2831097 \citep{Sodor} which was suspected to be the result of a light time
effect in a double stellar system. The large modulation
period and the lack of amplitude modulation exclude it to be attributed to a
Blazhko effect. But alternatively to the light time effect this phenomenon might
be due to a new modulation phenomenon with long period affecting at least RRc
stars.
\end{abstract}
\section{Introduction}
LINEAR 1169665 is a c sub-type RR Lyrae observed by TAROT \citep{tarot,klotz}
at Calern Observatory in the field of AE Leo which is routinely surveyed in the
frame of GEOS RR Lyr Survey (GRRS) \citep{leborgne2007,leborgne2012}. The star
was discovered by the survey Lincoln Near-Earth Asteroid Research
(LINEAR), a program searching for asteroids since 1998. \cite{Sesar2011}
have used LINEAR image database to search for time variable objects.
In particular, they published a general catalog of periodical variables
\citep{Palaversa} and a more complete catalog restricted to RR lyr stars
\citep{Sesar2013} to which LINEAR 1169665 belongs. \cite{Sesar2013} give a period
of 0.412982 days and an amplitude of 0.299 magnitude, magnitude at maximum being
14.432.
\begin{table}[ht]
\begin{center}
\begin{tabular}{lllllll}
\hline
                  &               &               &         &        &        \\
 star name        &  ra(J2000)    &  dec(J2000)   &   B     &   V    &  B-V   \\
                  &               &               &         &        &        \\
\hline
                  &               &               &         &        &        \\
 LINEAR 1169665   & 11:26:44.1168 & +17:51:11.556 &  14.474 & 14.426 &  0.048 \\
 UCAC4 538-052434 & 11:26:55.731  & +17:33:59.46  &  12.103 & 11.486 &  0.617 \\
                  &               &               &         &        &        \\
\hline
\end{tabular}
\end{center}
\caption{\label{tab1} Coordinates of LINEAR 1169665 and comparison star used for
reducing TAROT data. Magnitudes are those given by UCAC4 catalog \citep{Zacharias}.}
\end{table}
\section{Tarot observations and data mining}
\begin{figure}
\begin{center}
\begin{tabular}{p{9.5cm}p{6.5cm}}
\vspace{-2mm}\begin{center} \includegraphics[width=9cm]{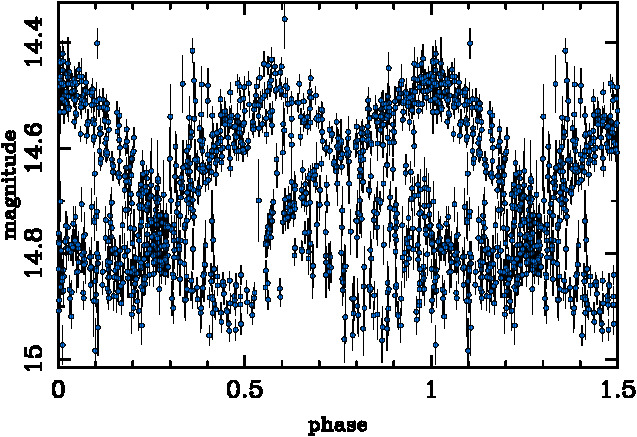} \end{center}&
\vspace{1mm} \captionof{figure}{\label{fig01}Folded light curve of TAROT data
using a period of 0.4129588 days. } \\
\end{tabular}
\end{center}
\end{figure}

TAROT CCD images of LINEAR 1169665 were obtained between JD 2453739 (2006
January 3) and 2457047 (2015 January 24). Photometry reduction of the images
used SExtractor software \citep{Bertin} to derive CV magnitudes of all stars
in the field. The images are obtained with no filter and are calibrated
with V magnitudes of comparison stars. Magnitudes of LINEAR 1169665 and comparison star UCAC4 538-052434
(table \ref{tab1}) were then extracted from SExtractor files. After selection on
internal signal to noise ratio of the TAROT images
it remains 1106 measurements covering a time interval of 9 years.\\
In order to check \cite{Sesar2013} period, we performed a frequency analysis on
TAROT data using \cite{Schwarzenberg-Czerny} algorithm. We found a period of
0.4129588 days similar to \cite{Sesar2013} result. Figure \ref{fig01} shows
the light curve of TAROT measurements folded with the period found and an epoch
of HJD 2456001.593.
One sees that the light curve appears to be bimodal which could be interpreted as
the result of a sudden period change during the observation time interval.
The measurements of individual maxima from TAROT data allow to explore this
hypothesis. Nine maxima were computed and are listed in table \ref{tab2} using
the elements HJD 2456001.593 + 0.4129588 E. \\
\begin{table}[ht]
\begin{center}
\begin{tabular}{lrrcc}
\hline
                        &             &        &              &                    \\
 Maximum HJD            &{\bf  O-C   }&    E   & Number of    & Duration of        \\
                        &{\bf (jous) }&        & measurements & observation (hours)\\
\hline
                        &             &        &              &                    \\
2456001.593 $\pm$0.010  &{\bf  0.000 }&    0   &    52        &   4.8              \\
2456006.5590$\pm$0.0097 &{\bf  0.010 }&   12   &    44        &   3.6              \\
2456288.636 $\pm$0.011  &{\bf  0.037 }&  695   &    36        &   3.8              \\
2456310.534 $\pm$0.010  &{\bf  0.048 }&  748   &    46        &   4.9              \\
2456327.4695$\pm$0.0097 &{\bf  0.052 }&  789   &    36        &   4.0              \\
2456367.507 $\pm$0.012  &{\bf  0.033 }&  886   &    50        &   4.8              \\
2456696.507 $\pm$0.014  &{\bf -0.096 }& 1683   &    41        &   5.3              \\
2456725.434 $\pm$0.014  &{\bf -0.076 }& 1753   &    50        &   4.6              \\
2456782.437 $\pm$0.013  &{\bf -0.061 }& 1891   &    42        &   4.1              \\
                        &             &        &              &                    \\
\hline
\end{tabular}
\end{center}
\caption{\label{tab2} Observed individual maxima of LINEAR 1169665 from TAROT data.}
\end{table}
The O-Cs of the maxima indeed show a sudden drop of about 4 hours (compared to
the period of 10 hours) between JD 2456367 (2013 March 15) and JD 2456696 (2014
February 7).\\
More information is available from survey public data. We collected CCD
measurements from ASAS-SN \citep{Shappee,Kochanek} from web site
\href{https://asas-sn.osu.edu/}{https://asas-sn.osu.edu/}
and from the discovery survey LINEAR. Concerning LINEAR a gzip-ed tar archive
containing star light curves may be downloaded from\\
\href{http://faculty.washington.edu/ivezic/linear/PaperIII/PLV.html}
{http://faculty.washington.edu/ivezic/linear/PaperIII/PLV.html}.
LINEAR data cover a time interval from JD 2452614 (2002 December 5) to JD 2454619
(2008 June 1$^{st}$) and ASAS-SN data from JD 2455969 (2012 February 11) to JD
2458486 (2019 January 2). The three data sets are then complementary
to cover a time interval of 17 years from December 2002 to January 2019. \\
Individual maxima are hardly obtained from LINEAR and ASAS-SN data because both
collect few measurements each night. On the other hand, the light curves of RRc
stars show a flat maximum, and even sometimes a double maximum: it is then better
to define mean maxima determined on folded light curves from given time intervals.
Given the quite small number of measurements (table \ref{tab3}) of each surveys,
we built folded light curves on intervals of one year (see light curves in appendix).
Table \ref{tab3} lists the mean maxima obtained in such a way.
Note that 2006 LINEAR data before November are not compatible with 2005 and 2007
data due to period change. After November 2006, data are compatible with 2007 data.
However, there are no measurements at maximum in 2006. In 2012 and 2013, there
are not enough measurements in ASAS-SN data to estimate a mean maximum. \\
Put together, the 19 maxima from the three surveys suggested a period slightly
longer than the one determined by the spectral analysis of TAROT data. We
determined elements which satisfy all the maxima as a mean, with however an O-C
amplitude of $\pm$1h48m:
\begin{equation} \label{eq:elem} 2456001.568\ +\ 0.412963\ E\end{equation}
\begin{table}[ht]
\begin{center}
\begin{tabular}{lrl}
\hline
          &              &                       \\
   year   & Number of    & Mean maximum          \\
          & measurements & HJD                   \\
\hline
          &              &                       \\
\multicolumn{3}{c}{\bf LINEAR}     \\
2002-2003 &     108      & 2452962.962$\pm$0.005 \\
2004-2005 &      75      & 2453062.849$\pm$0.005 \\
2006      &      50      &                       \\
2007      &      20      & 2454154.792$\pm$0.004 \\
2008      &      52      & 2454539.735$\pm$0.004 \\
          &              &                       \\
\multicolumn{3}{c}{\bf ASAS-SN}     \\
2012-2013 &      68      &                       \\
2014      &      86      & 2456747.334$\pm$0.02  \\
2015      &     134      & 2457112.386$\pm$0.02  \\
2016      &     180      & 2457478.304$\pm$0.02  \\
2017      &     254      & 2457843.438$\pm$0.02  \\
2018      &     130      & 2458208.471$\pm$0.02  \\
          &              &                       \\
\hline
\end{tabular}
\end{center}
\caption{\label{tab3} Observed mean maxima of LINEAR 1169665 from LINEAR and ASAS-SN surveys.}
\end{table}
\begin{figure}
\begin{center}
\begin{tabular}{p{9.5cm}p{6.5cm}}
\vspace{-9mm}\includegraphics[width=8cm]{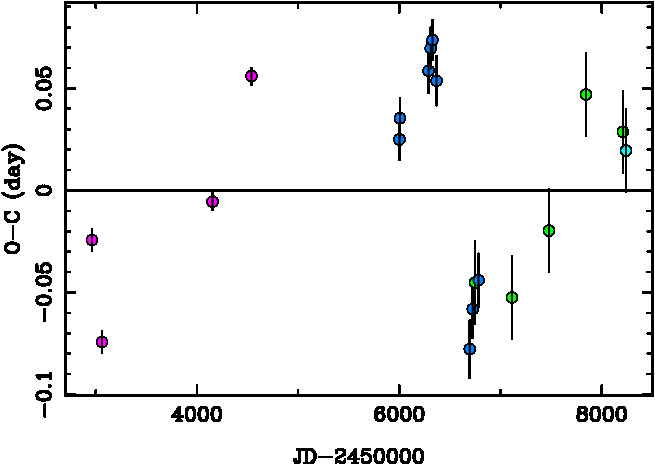}&
\vspace{1mm} \captionof{figure}{\label{fig02} O-C curve of LINEAR 1169665 maxima.
TAROT data correspond to blue points. Purple points are LINEAR data. ASAS-SN data
are represented by green points for V filter and turquoise for g filter.}
\end{tabular}
\end{center}
\end{figure}
\section{Period modulation}
\begin{figure}
\begin{center}
\begin{tabular}{p{8cm}p{8cm}}
\vspace{-9mm}\begin{center} \includegraphics[width=8cm]{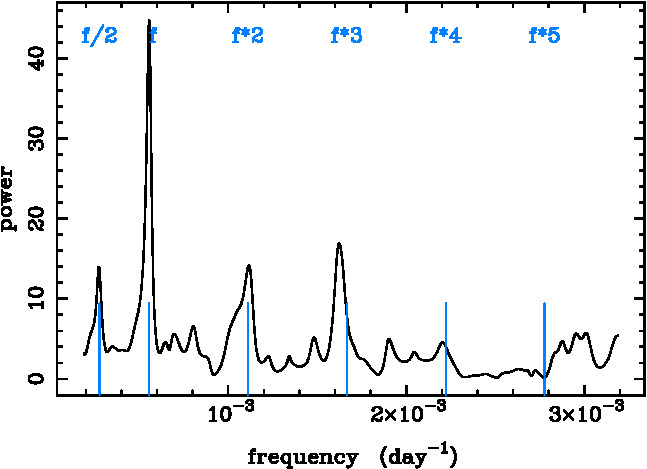} \end{center}&
\vspace{1mm} \captionof{figure}{\label{fig03} Periodogram plot for O-C periodicity.
The value of main frequency $f$ is 5.5552 10$^{-4}$ d$^{-1}$ corresponding to
a period of 1800.1 days. } \\
\vspace{-9mm}\begin{center} \includegraphics[width=8cm]{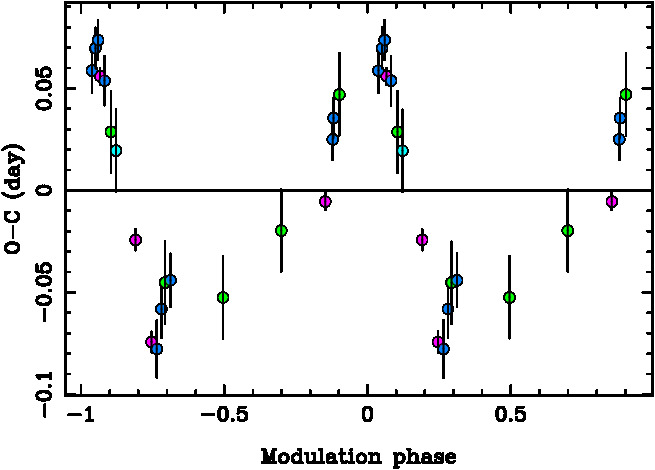} \end{center}&
\vspace{-9mm}\begin{center} \includegraphics[width=8cm]{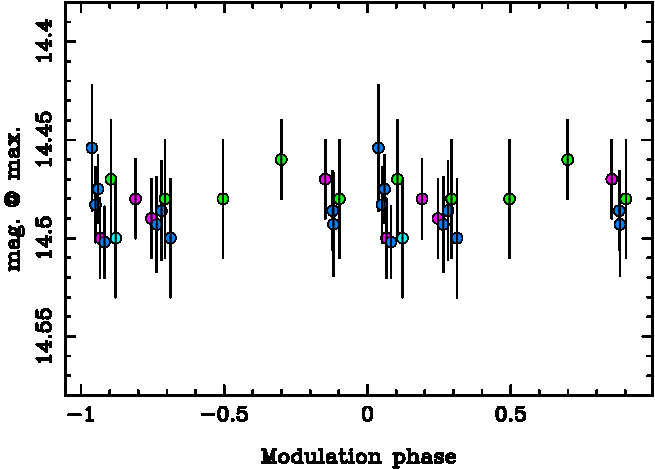} \end{center}\\
\end{tabular}
\captionof{figure}{\label{fig04} O-C curve and magnitude at maximum folded on
modulation period of 1800.1 days. Color codes are the same as for figure \ref{fig02}}
\end{center}
\end{figure}
The O-C curve using elements (\ref{eq:elem}) is displayed in figure \ref{fig02}.
The dispersion is large but we are tempted to see a periodicity in the O-C variation.
We then performed a frequency analysis (\cite{Schwarzenberg-Czerny} algorithm)
and found a period of 1800.1 days as shown in figure \ref{fig03} which confirm the
visual impression. \\
\begin{figure}
\begin{center}
\begin{tabular}{p{8cm}p{8cm}}
\vspace{-9mm}\begin{center} \includegraphics[width=8cm]{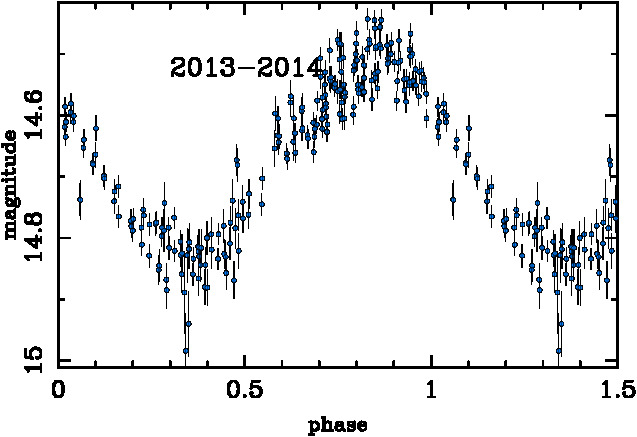} \end{center}&
\vspace{-9mm}\begin{center} \includegraphics[width=8cm]{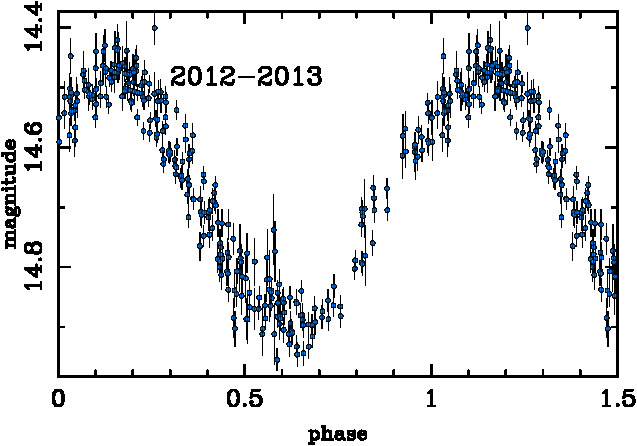} \end{center}\\
\end{tabular}
\captionof{figure}{\label{fig_tarot} Folded light curve of TAROT data for two
seasons, 2012-2013 and 2013-2014. The phases are computed with elements
(\ref{eq:elem}) }
\end{center}
\end{figure}
Table \ref{tab4} displays a summary of the maxima measured from the data of TAROT,
LINEAR and ASAS-SN surveys. The O-C are calculated with the elements (\ref{eq:elem})
and are those plotted in figure \ref{fig02}. The magnitudes at maximum are
listed in the third column. Tarot and LINEAR give magnitudes in CV photometric
filter. ASAS-SN images are filtered and magnitudes
are given for V filter with the exception of the last maximum which was measured
on g filter data.
The magnitudes of LINEAR and ASAS-SN surveys are quite similar, even if the last
maximum was obtained with g filter, bluer than V filter. Tarot magnitudes,
however, seem shifted relative to the other surveys. This appears clearly in the
mean maximum magnitude: \\
- TAROT: 9 max., mean maximum mag. 14.486, standard deviation 0.0146 mag. \\
- LINEAR and ASAS-SN: 10 max., mean maximum mag. 14.401, standard dev.
0.0129 mag. \\
Unlike O-C, magnitude at maximum is constant over the 17 years of observation,
within measurement precision.
The difference of 0.081 magnitudes between TAROT and the other series may be
due to the calibration method: a single comparison star for TAROT and ensemble
calibration for the other series.
O-C curve and magnitude at maximum folded on modulation period of 1800.1 days
are plotted in figure \ref{fig04}. The origin time used is 2456220. This value
was chosen as an estimation of the maximum of O-Cs. The amplitude of the  O-C
modulation is 0.15 days. We subtracted 0.081 magnitudes from magnitude at maximum
of the TAROT data in order to equal the mean values of the 3 series. \\
\begin{table}[ht]
\begin{center}
\begin{tabular}{lrll}
\hline
                        &         &                  &         \\
    Maximum             &    O-C  & Magnitude at     & Survey  \\
                        &  (days) &   maximum        &         \\
                        &         &                  &         \\
2452962.962$\pm$0.005   & -0.0242 & 14.40$\pm$0.02   & LINEAR  \\
2453062.849$\pm$0.005   & -0.0743 & 14.41$\pm$0.02   & LINEAR  \\
2454154.792$\pm$0.004   & -0.0055 & 14.39$\pm$0.02   & LINEAR  \\
2454539.735$\pm$0.004   &  0.0560 & 14.42$\pm$0.02   & LINEAR  \\
2456001.593$\pm$0.01    &  0.0250 & 14.486$\pm$0.02  &  Tarot  \\
2456006.559$\pm$0.0097  &  0.0354 & 14.493$\pm$0.026 &  Tarot  \\
2456288.636$\pm$0.011   &  0.0587 & 14.454$\pm$0.032 &  Tarot  \\
2456310.534$\pm$0.01    &  0.0697 & 14.483$\pm$0.019 &  Tarot  \\
2456327.4695$\pm$0.0097 &  0.0737 & 14.475$\pm$0.017 &  Tarot  \\
2456367.507$\pm$0.012   &  0.0538 & 14.502$\pm$0.018 &  Tarot  \\
2456696.507$\pm$0.014   & -0.0777 & 14.493$\pm$0.024 &  Tarot  \\
2456725.434$\pm$0.014   & -0.0581 & 14.486$\pm$0.025 &  Tarot  \\
2456747.334$\pm$0.02    & -0.0452 & 14.40$\pm$0.03   & ASAS-SN \\
2456782.437$\pm$0.013   & -0.0440 & 14.500$\pm$0.03  &  Tarot  \\
2457112.386$\pm$0.02    & -0.0525 & 14.40$\pm$0.03   & ASAS-SN \\
2457478.304$\pm$0.02    & -0.0197 & 14.38$\pm$0.02   & ASAS-SN \\
2457843.43 $\pm$0.02    &  0.0470 & 14.40$\pm$0.03   & ASAS-SN \\
2458208.471$\pm$0.02    &  0.0287 & 14.39$\pm$0.03   & ASAS-SN \\
2458239.434$\pm$0.02    &  0.0195 & 14.42$\pm$0.03   & ASAS-SN; g filter \\
                        &         &                  &         \\
\hline
\end{tabular}
\end{center}
\caption{\label{tab4} Summary of observed maxima of LINEAR 1169665.}
\end{table}
Figure \ref{fig_tarot} shows a sample of folded light curves of TAROT data for
two seasons, 2012-2013 and 2013-2014. The other seasons do not have complete
light curves. The phases are computed with elements (\ref{eq:elem}). Note the
phases of the maxima corresponding to the O-C seen in figure \ref{fig02} and
\ref{fig04}. These two seasons are before and after the 2013 sudden drop of O-Cs.
\section{Interpretation and conclusion}
Analyzing measurements of the RRc LINEAR 1169665 made by telescope TAROT in
Calern from 2006 to 2015 and by the surveys LINEAR (2002-2008) and ASAS-SN
(2012-2019), we have found that this star exhibits a period modulation with a
period of 1800.1 days. No modulation is found in the magnitude at maximum.
Because the period of the modulation is very long and the magnitude at maximum
shows no modulation, we can say it is hardly a Blazhko effect. The behavior of
LINEAR 1169665 looks like the one of the RRc star KIC 2831097 in many aspects.
\citep{Sodor} have shown that KIC 2831097 observed by KEPLER satellite have
period modulation with a period of 753 days with no amplitude modulation. The
amplitude of the variation of O-C is 0.036 days. \\
Logically, this behavior can be interpreted as a light time effect (LiTE) in a
double stellar system. In the case of KIC 2831097, \citep{Sodor} translated the
O-C amplitude as the time necessary to the light to travel across the orbit.
They found as projected semi-major axis $a\ sin\ i=3.143 AU$, where $i$ is the
obliquity of KIC 2831097 orbit on the line of sight.
They also computed that the companion should have a mass of 8.4 M$_{\odot}$
assuming the RR Lyr has a mass of 0.6 M$_{\odot}$. For LINEAR 1169665,
the binarity hypothesis interprets the modulation period of 1800.1 days as the
orbital period and the amplitude of the O-C of 0.15 days as the light travel time
 across
the orbit corresponding to a projected orbit size of 13 AU. Given the shape of
the O-C folded curve, similar to KIC 2831097, the ellipticity of the orbit
should be large. In such case, the light travel time we calculate does not
correspond the projection of the major axis but to another cord, necessarily
smaller.
\\

Another interpretation is that we are facing to a new modulation type affecting
RRc (at least) differing from Blazhko effect with a very long period and no amplitude
modulation.
\section{Acknowledgements}
This research has made use of the SIMBAD database, operated at CDS, Strasbourg, France
\citep{Wenger} and of the VizieR catalogue access tool, CDS, Strasbourg,
France. The original description of the VizieR service was published in
A\&AS 143, 23, http://vizier.u-strasbg.fr/viz-bin/VizieR\\
This research also has made use of the International Variable Star Index (VSX)
database, operated at AAVSO, Cambridge, Massachusetts, USA.
https://www.aavso.org/vsx/index.php
and of the GEOS database of RR Lyr stars \citep{leborgne2007} hosted
at Institut de Recherche en Astrophysique et Plan\'etologie, Toulouse, France
\href{http://rr-lyr.irap.omp.eu/dbrr/}{http://rr-lyr.irap.omp.eu/dbrr/} \\
The authors thank the referee for his useful comments.

\newpage
\appendix
\section{Appendix: Mean light curves}
\begin{figure}[ht]
\begin{center}
\begin{tabular}{p{6.9cm}p{6.9cm}}
\vspace{-9mm}\begin{center} \includegraphics[width=6.9cm]{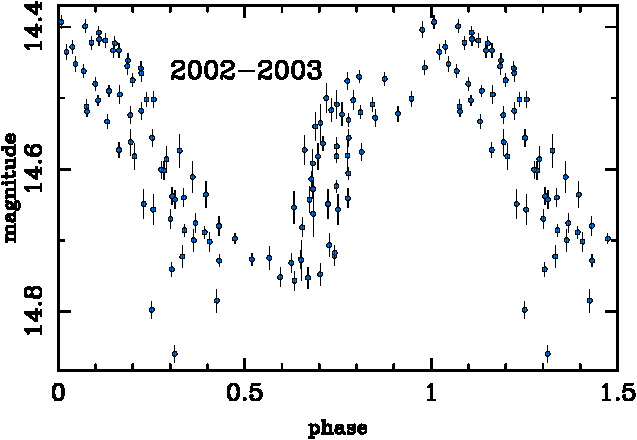} \end{center}&
\vspace{-9mm}\begin{center} \includegraphics[width=6.9cm]{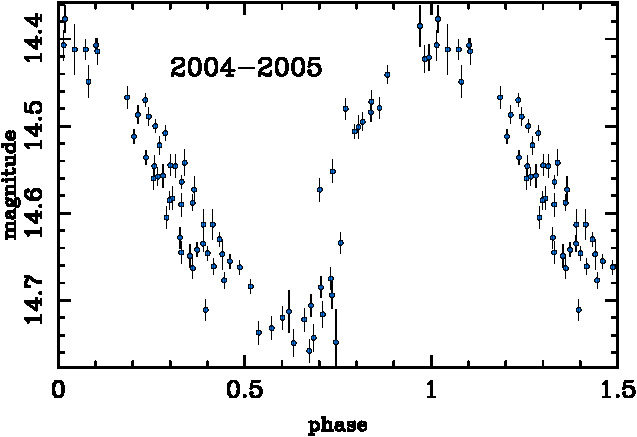} \end{center}\\
\vspace{-9mm}\begin{center} \includegraphics[width=6.9cm]{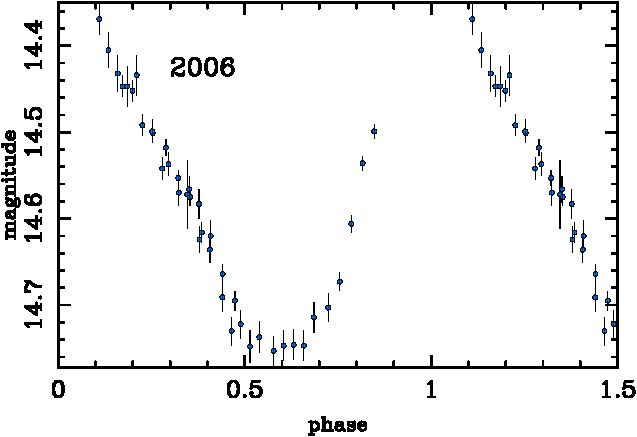} \end{center}&
\vspace{-9mm}\begin{center} \includegraphics[width=6.9cm]{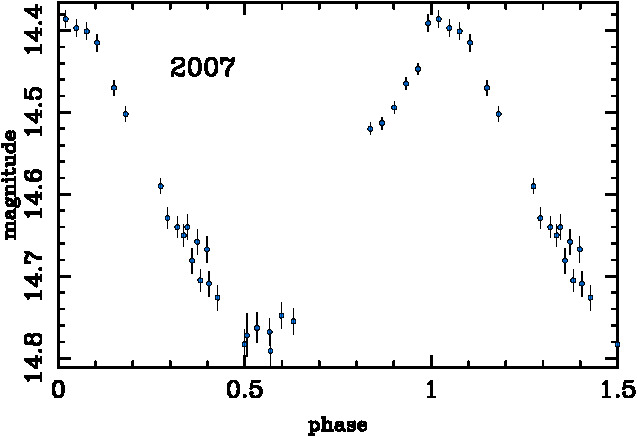} \end{center}\\
\vspace{-9mm}\begin{center} \includegraphics[width=6.9cm]{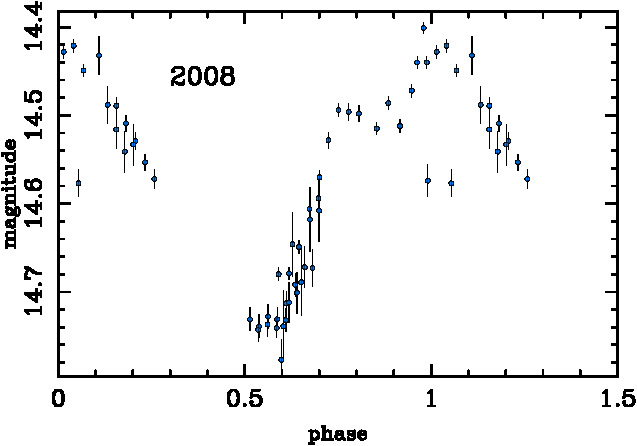} \end{center}&
\vspace{-9mm}\begin{center}
\captionof{figure}
{\label{fig05} Folded light curves of LINEAR data for LINEAR 1169665 from 2002 to 2008.}
\end{center}\\
\end{tabular}
\end{center}
\end{figure}

\begin{figure}[ht]
\begin{center}
\begin{tabular}{p{6.9cm}p{6.9cm}}
\vspace{-9mm}\begin{center} \includegraphics[width=6.9cm]{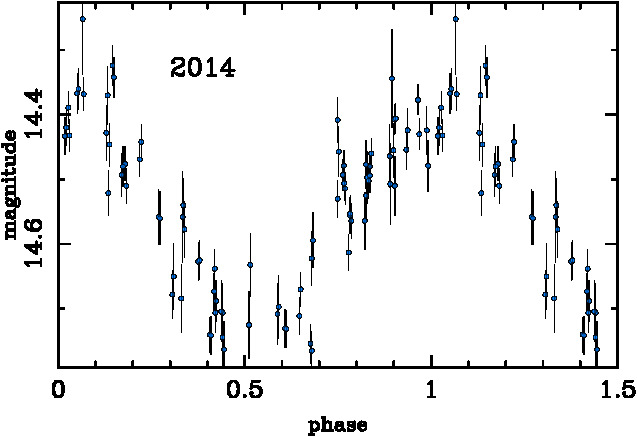} \end{center}&
\vspace{-9mm}\begin{center} \includegraphics[width=6.9cm]{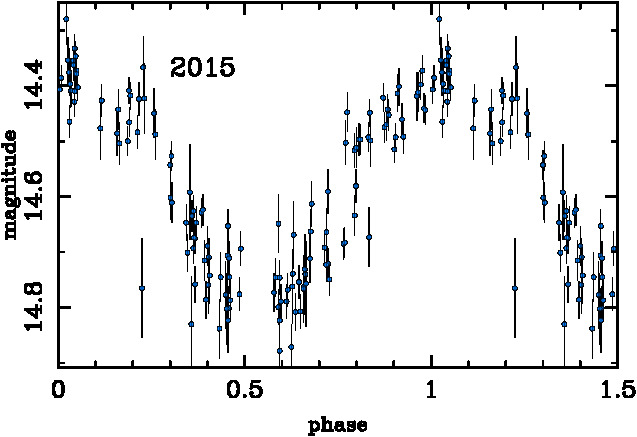} \end{center}\\
\vspace{-9mm}\begin{center} \includegraphics[width=6.9cm]{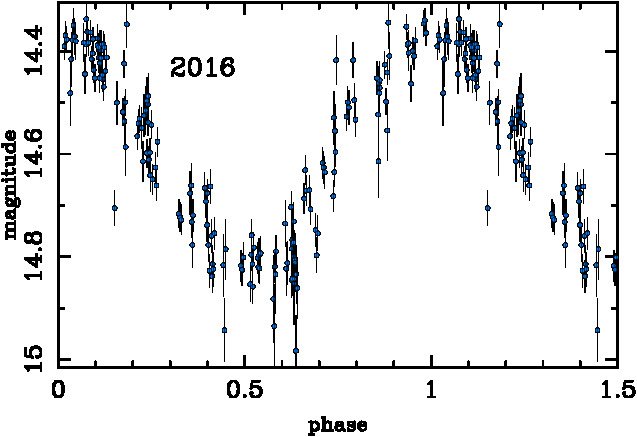} \end{center}&
\vspace{-9mm}\begin{center} \includegraphics[width=6.9cm]{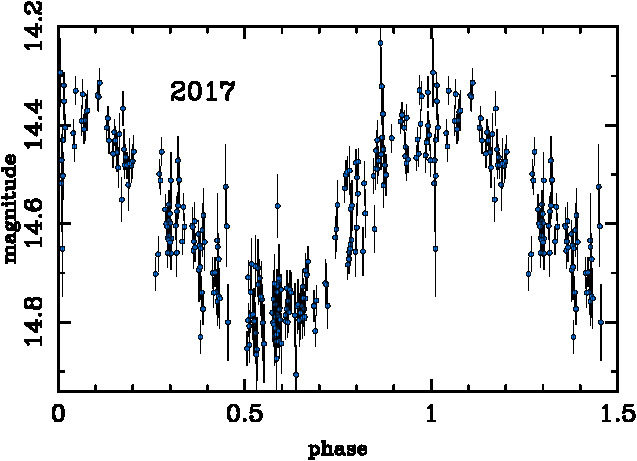} \end{center}\\
\vspace{-9mm}\begin{center} \includegraphics[width=6.9cm]{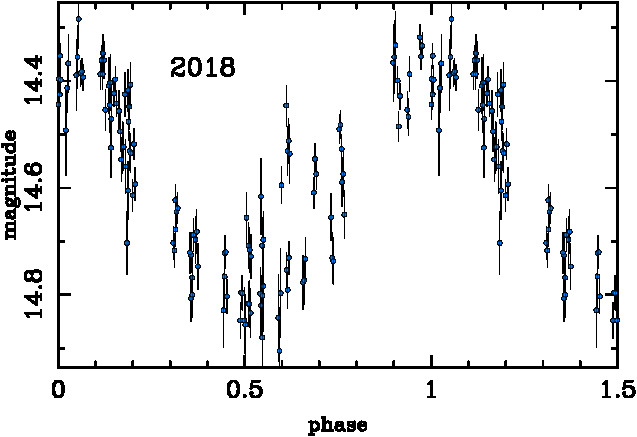} \end{center}&
\vspace{-9mm}\begin{center} \includegraphics[width=6.9cm]{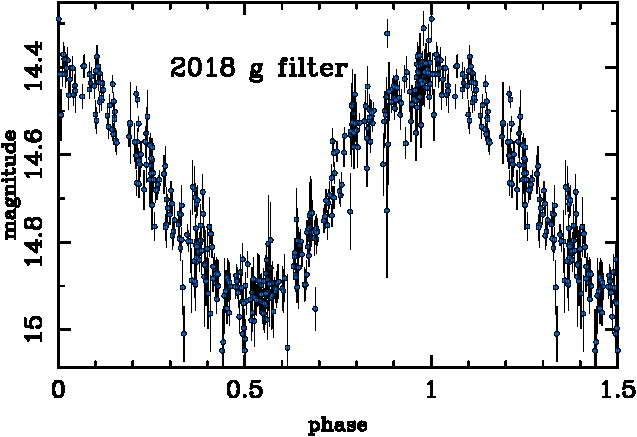} \end{center}\\
\end{tabular}
\captionof{figure}
{\label{fig06} Folded light curves of ASAS-SN data for LINEAR 1169665 from 2012 to 2018.
Filter used is V, except for the last curve which is obtained with g filter.}
\end{center}
\end{figure}

The figures \ref{fig05} and \ref{fig06} show the yearly folded light curves of
LINEAR 1169665 from LINEAR and ASAS-SN surveys used to compute mean maxima.
For each curve, the phase origin corresponds to the mean maximum as given in
table \ref{tab3}, with the exception of 2006 LINEAR curve. \\
The comparison of figures \ref{fig_tarot} to \ref{fig06} gives the impression
that the hump before maximum is more pronounced on some folded light curves, for
example, LINEAR 2002-2003 and 2008. If real, it is difficult to say with the
present data if this is connected to the period modulation or to the difference
of detector spectral response. \\

\end{document}